\input harvmac
\input psfig
\newcount\figno
\figno=0
\def\fig#1#2#3{
\par\begingroup\parindent=0pt\leftskip=1cm\rightskip=1cm\parindent=0pt
\global\advance\figno by 1
\midinsert
\epsfxsize=#3
\centerline{\epsfbox{#2}}
\vskip 12pt
{\bf Fig. \the\figno:} #1\par
\endinsert\endgroup\par
}
\def\figlabel#1{\xdef#1{\the\figno}}
\def\encadremath#1{\vbox{\hrule\hbox{\vrule\kern8pt\vbox{\kern8pt
\hbox{$\displaystyle #1$}\kern8pt}
\kern8pt\vrule}\hrule}}

\overfullrule=0pt

%
\def\underarrow#1{\vbox{\ialign{##\crcr$\hfil\displaystyle
 {#1}\hfil$\crcr\noalign{\kern1pt\nointerlineskip}$\longrightarrow$\crcr}}}
%
\def\tilde{\widetilde}
\def\bar{\overline}

\def\inbar{\vrule height1.5ex width.4pt depth0pt}
\def\IC{\relax\hbox{\kern.25em$\inbar\kern-.3em{\rm C}$}}
\def\IR{\relax\hbox{\kern.25em$\inbar\kern-.3em{\rm R}$}}
\def\IZ{\relax\ifmmode\hbox{Z\kern-.4em Z}\else{Z\kern-.4em Z}\fi}

\font\zfont = cmss10 

\def\bigone{\hbox{1\kern -.23em {\rm l}}}
\def\ZZ{\hbox{\zfont Z\kern-.4emZ}}


\def\drawbox#1#2{\hrule height#2pt
        \hbox{\vrule width#2pt height#1pt \kern#1pt
              \vrule width#2pt}
              \hrule height#2pt}

\def\Fund#1#2{\vcenter{\vbox{\drawbox{#1}{#2}}}}
\def\Asym#1#2{\vcenter{\vbox{\drawbox{#1}{#2}
              \kern-#2pt       
              \drawbox{#1}{#2}}}}

\def\fund{\Fund{6.5}{0.4}}

\batchmode
  \font\bbbfont=msbm10
\errorstopmode
\newif\ifamsf\amsftrue
\ifx\bbbfont\nullfont
  \amsffalse
\fi
\ifamsf
\def\IR{\hbox{\bbbfont R}}
\def\IC{\hbox{\bbbfont C}}

\def\IZ{\hbox{\bbbfont Z}}

\def\antifund{\bar{\fund}}
\midinsert
\endinsert


\nref\polchinski{ J. Polchinski, ``TASI Lectures on D-branes'',in
{\it Fields Strings and Duality} TASI-96, Eds. C. Efthimiou and B. Greene, World
Scientific (1997), hep-th/9611050.}

\nref\giveon{A. Giveon and D. Kutasov, ``Brane Dynamics and Gauge Theory'',
hep-th/9802067.}

\nref\ghm{ M. Green, J.A. Harvey and G. Moore, ``I-Brane Inflow and Anomalous Couplings
on D-Branes'', Class. Quant. Grav. {\bf 14} (1997) 47, hep-th/9605033.}

\nref\cheung{Y-K. E. Cheung and Z. Yin, ``Anomalies, Branes and Currents'', Nucl. Phys.
B {\bf 517} (1998) 69, hep-th/9710206.}

\nref\moore{ R. Minasian and G. Moore, ``K-Theory and Ramond-Ramond Charge'',
JHEP {\bf 9711} (1997) 002, hep-th/9710230.}

\nref\strass{ M. Strassler, ``On Phases of Gauge Theories and the Role of Non-BPS
Solitons in Field Theory'', hep-th/9808073.}

\nref\sen{A. Sen, ``Stable Non-BPS States In String Theory'',  JHEP {\bf 9806} (1998) 007,
hep-th/9803194;
``Stable Non-BPS Bound States of BPS $D$-Branes'' JHEP {\bf 9808} (1998) 010,
hep-th/9805019; ``Tachyon
Condensation on the Brane-Antibrane System'', JHEP {\bf 9808} (1998) 012, hep-th/989805170.}

\nref\senuno{ A. Sen, ``SO$(32)$ Spinors of Type I and Other Solitons on Brane-Antibrane
Pair'',  JHEP {\bf 9809} (1998) 023, hep-th/9808141.}

\nref\sendos{A. Sen, ``Type I $D$-Particle and Its Interactions'', JHEP {\bf 9810} (1998)
021, hep-th/9809111.}

\nref\senew{ A. Sen, ``BPS D-branes on Non-supersymmetric Cycles'', JHEP {\bf 9812} (1998)
021, hep-th/9812031.}

\nref\bergmanone{O. Bergman and M.R. Gaberdiel, ``Stable non-BPS D-particles'',
Phys. Lett. B {\bf 441} (1998) 133, hep-th/9806155.}

\nref\bergmantwo{O. Bergman, ``Stable non-BPS Dyons in ${\cal N}=2$ SYM'',
JHEP {\bf 9905} (1999) 004, hep-th/9811064.}

\nref\witten{E. Witten, ``$D$-Branes and K-Theory'', JHEP {\bf 9812} (1998) 019,
hep-th/9810188.}

\nref\atiyah{M.F. Atiyah, {\it K-Theory} (W.A. Benjamin, New York, 1967).}

\nref\bott{R. Bott, {\it Lectures on K(X)} (W.A. Benjamin, New York, 1969).}

\nref\karo{M. Karoubi, {\it K-Theory, An Introduction} Springer-Verlag (1978);
{\it K-Th\'eorie} Les Presses de L'Universit\'e de
Montr\'eal (1971).}

\nref\spin{ H.B. Lawson, Jr. and M-L. Michelsohn, {\it Spin Geometry} (Princeton University Press,
Princeton, 1989).}

\nref\kbranes{P. Ho$\check{r}$ava, ``Equivariant Topological Sigma Models'',  Nucl. Phys. B
{\bf 418} (1994) 571; M. Bershadsky, V. Sadov and C. Vafa, ``D-branes and Topological Field Theories'',
hep-th/9511222; A. Strominger, ``Open $p$-branes'', hep-th/9512059.}

\nref\as{M.F. Atiyah and G.B. Segal, ``Equivariant K-Theory'', Lectures Notes University
of Warwick, (1965); ``Equivariant K-Theory and Completion'',
J. Diff. Geom. {\bf 3} (1969) 1-18.}

\nref\segal{G.B. Segal, ``Equivariant K-Theory'', IHES Sci. Publ. Math. No. 34 (1968) 129-151.}

\nref\greenless{ J.P.C. Greenless, ``An Introduction to Equivariant K-theory'', in {\it Equivariant
Homotopy and Cohomology} Ed. J.P. May, Conference Board of the Mathematical Sciences {\bf 91} (1996) pp.
143-152.}

\nref\dm{M.R. Douglas and G. Moore, ``D-branes, Quivers and ALE
Instantons, hep-th/9603167.}

\nref\johnson{C.~V.~Johnson, R.~C.~Myers, ``Aspects of Type IIB Theory on
ALE Spaces'', Phys. Rev. D {\bf 55} (1997) 6382, hep-th/9610140.}

\nref\six{K.~Intriligator, ``RG Fixed Points in Six Dimensions
Via Branes at Orbifold Singularities'', Nucl. Phys. B {\bf 496} (1997) 177,
hep-th/9702038, ``New String Theories In Six Dimensions Via Branes at
Orbifolds Singularities'', hep-th/9708117.}

\nref\bi{
 J.~Blum, K.~Intriligator, ``Consistency Conditions for
Branes at Orbifold Singularities'', Nucl. Phys. B {\bf 506} (1997) 223,
hep-th/9705030; ``New Phases of String Theory and 6d RG Fixed Points Via
Branes at Orbifold Singularities'', Nucl. Phys. B {\bf 506} (1997) 199,
hep-th/9705044.}

\nref\dgm{ M.R. Douglas, B.R. Greene and D.R. Morrison, ``Orbifolds
Resolution by D-branes'', Nucl. Phys. B {\bf 506} (1997) 84, hep-th/9704151.}

\nref\mohri{ K. Mohri, ``D-branes and Quotients Singularities of Calabi-Yau
Fourfolds'', Nucl. Phys B {\bf 521} (1998)161, hep-th/9707012; ``Kahler Moduli Space of a
D-brane at Orbifold Singularities'',  Commun. Math. Phys. {\bf 202} (1999) 669,
hep-th/9806052.}

\nref\k{ P. Kronheimer, ``The Construction of ALE Spaces as Hyper-K\"ahler Quotients'',
J. Diff. Geom. {\bf 28} (1989) 665; ``A Torreli-Type Theorem For Gravitational
Instantons'', J. Diff. Geom. {\bf 29} (1989) 685.}

\nref\kn{P. Kronheimer and H. Nakajima, ``Yang-Mills Instantons on ALE Gravitational
Instantons'', Math. Ann. {\bf 288} (1990) 263.}

\nref\seiberg{N. Seiberg, ``Non-trivial Fixed Points of the Renormalization Group in Six Dimensions'',
Phys. Lett. B {\bf 390} (1997) 169, hep-th/9609161.}

\nref\julie{J.D. Blum, ``Anomaly Inflow at Singularities'', Phys. Lett. B {\bf 435} (1998) 319,
hep-th/9806012.}

\nref\sma{M.F. Atiyah, ``Bott Periodicity and the Index of Elliptic Operators'', Quart. J. Math.
Oxford {\bf 19} (1968) 113.}

\nref\small{E. Witten, ``Small Instantons in String Theory'', Nucl. Phys. B {\bf 460} (1996)
541,  hep-th/9511030.}

\nref\douglas{ M.R. Douglas, ``Branes within Branes'', hep-th/9512077.}

\nref\ast{ M.F. Atiyah and I.M. Singer, ``The Index of Elliptic Operators III'', Ann. of Math.
{\bf 87} (1968) 546.}

\nref\aps{ M.F. Atiyah, V.K. Patodi and I.M. Singer, ``Spectral Asymmetry and Riemannian
Geometry, I, Math. Proc. Camb. Phil. Soc. {\bf 77} (1975) 43; II, {\bf 78} (1975) 405;
III, {\bf 79} (1976) 71.}

\nref\pope{G.W. Gibbons, C.N. Pope and H. R\"omer, Nucl. Phys. B {\bf 157} (1979) 377.}

\nref\ads{J. Maldacena,  ``The Large $N$ Limit of Superconformal Field Theories and
Supergravity'', hep-th/9711200; S.S. Gubser, I.R. Klebanov and A.M. Polyakov, ``Gauge Theory
Correlators From
Non-critical String Theory'', Phys. Lett. B {\bf 428} (1998) 105, hep-th/9802109;
E. Witten, ``Anti-de Sitter Space and Holography'', Adv. Theor. Math. Phys. {\bf 2} (1998)
253, hep-th/9802150.}

\nref\ks{S. Kachru and E. Silverstein, ``4d Conformal Field Theories and
Strings on Orbifolds'', Phys. Rev. Lett. {\bf 80} (1998) 4855,
hep-th/9802183.}

\nref\vafa{ A. Lawrence, N. Nekrasov and C. Vafa, ``On Conformal Field Theories
in Four Dimensions'', Nucl. Phys. B {\bf 533} (1998) 199, hep-th/9803015.}

\nref\hu{A. Hanany and A.M. Uranga, ``Brane Boxes and Branes on
Singularities'', JHEP {\bf 9805} (1998) 013, hep-th/9805139.}

\nref\angel{H. Garc\'{\i}a-Compe\'an and A.M. Uranga, ``Brane Box Realization of Chiral Gauge Theories in
Two Dimensions'',  Nucl. Phys. B {\bf 539} (1999) 329, hep-th/9806177.}

\nref\kw{I.R. Klebanov and E. Witten, ``Superconformal Field Theory at a Calabi-Yau Singularity'',
Nucl. Phys. B {\bf 536} (1998) 199, hep-th/9807080.}

\nref\conifold{D.R. Morrison and M.R. Plasser, ``Non-Spherical Horizons, I'',
hep-th/9810201; K. Oh and R. Tatar, ``Three-dimensional SCFT From M2 Branes at Conifold Singularities'',
hep-th/9810244; A.M. Uranga, ``Brane Configurations for Branes at Conifolds'',
JHEP {\bf 9901} (1999) 022, hep-th/9811004;
S. Gukov, M. Rangamani and E. Witten, ``Dibaryons, Strings and Branes in AdS Orbifold Models'',
hep-th/9811048; K. Dasgupta and S. Mukhi, ``Brane Constructions, Conifolds and M-Theory'',
JHEP {\bf 9812} (1998) 025, hep-th/9811139; E. Lopez, ``A Family of ${\cal N}=1$ SU$(N)^k$ Theories From Branes at Orbifolds'',
JHEP {\bf 9902} (1999) 019, hep-th/9812025; R. de Mello Koch, K. Oh and R. Tatar, ``Moduli Space For Conifolds as Intersection
of Orthogonal D6 Branes'', hep-th/9812097.}

\nref\horava{P. Ho$\check{r}$ava, ``Type IIA D-Branes, K-Theory, and Matrix Theory'',
Adv. Theor. Math. Phys. {\bf 2} (1998) 1373, hep-th/9812135.}

\nref\orbi{E. Witten, ``Fivebranes and $M$-Theory on an Orbifold'', Nucl. Phys. B {\bf 463} (1996)
383, hep-th/9512219.}


\Title{hep-th/9812226, IASSNS-HEP-98/106}
{\vbox{\centerline{D-branes in Orbifold Singularities}
\medskip
\centerline{and Equivariant K-Theory  }}}
\smallskip
\centerline{Hugo
Garc\'{\i}a-Compe\'an\foot{Also {\it Departamento de
F\'{\i}sica, Centro de Investigaci\'on y de Estudios Avanzados del IPN,
Apdo. Postal 14-740, 07000, M\'exico D.F., M\'exico.} E-mail:
compean@sns.ias.edu}}
\smallskip
\centerline{\it School of Natural Sciences}
\centerline{\it Institute for Advanced Study}
\centerline{\it Olden Lane, Princeton, NJ 08540, USA}

\bigskip
\medskip
\vskip 1truecm
\noindent

The study of brane-antibrane configurations in string theory leads to the
understanding of supersymmetric D$p$-branes as the bound states of higher
dimensional branes. Configurations of pairs brane-antibrane do admit in a
natural way their description in terms of K-theory. We analyze
configurations of brane-antibrane at fixed point orbifold singularities in
terms of equivariant K-theory as recently suggested by Witten. Type I and
IIB fivebranes and small instantons on ALE singularities are described in
K-theoretic terms and their relation to Kronheimer-Nakajima construction
of instantons is also provided.  Finally the D-brane charge formula is
reexamined in this context.


\noindent

\Date{December, 1998}



\newsec{Introduction}

D-branes are extremely interesting objects in string theory and thus their study is still
matter of vigorous research (for a review see \polchinski). Many surprises
have been found in the study of various physical situations of D-branes
and  surely many other surprises remain to be discovered in the near future.
Among their multiple
applications D-branes have been used to study the strong coupling dynamics of
supersymmetric field theories in various dimensions through the construction of
configurations of intersecting branes (for a review see \giveon).
Gauge theories in $p+1$ dimensions with
sixteen supercharges can be obtained as the world-volume theories of flat
infinite D$p$-branes.
Intersecting brane configurations are described by BPS
bound states of the corresponding superstring theory where the D-brane is
living. Intersection of D-branes generically has
zero fermionic modes which are chiral and therefore anomalous on the branes.
D-branes carries RR charges and the definition of this charge, may be given
 through the implementation of
this anomaly cancellation via the inflow mechanism from the bulk where the D-branes are living
\ghm. The consideration of these facts in topological terms and a further
generalization  to non-trivial normal bundle to the cycles
 where the D-branes are wrapped lead to a
formula for the brane charge which widely suggest that it takes values not in the
homology of spacetime but in the K-theory of the spacetime \refs{\cheung,\moore}.

Recently some works have been made in the context of stable
non-supersymmetric states (non-BPS states) in field theory \strass\ and string theory
\refs{\sen,\senuno,\sendos,\senew,\bergmanone,\bergmantwo}. In particular, in string theory context,
Sen have shed light on the structure of stable non-BPS states in type I and II
superstring theories by constructing non-BPS states through the consideration
of brane-antibrane pairs and their dynamical properties \senuno. Boundary state analysis
of these configurations leads to a world-volume field theory, on the brane
pair, without the usual super Maxwell multiplet but with a tachyon field on the
world-volume which survives the GSO projection.
Potential of the tachyon field of this system is shown to be of the same form as the potential for a
kink-anti-kink topological defect which interpolates between the two minima of the tachyonic
potential. This system is already not supersymmetric but it has an stable vacuum. The vacuum
configuration is reached through the mechanism of tachyon condensation.
Generalization to other higher dimensional topological defects  was considered
as well in \senuno.

Following Sen's results very recently Witten \witten, classified the non-super\-sym\-metric
brane configurations in terms of the mathematical structure known as topological K-theory
(for some reviews see \refs{\atiyah,\bott,\karo,\spin}). Some earlier evidence of the possible relation
between D-branes and K-theory can be found in \kbranes.
At the same paper
Witten found further evidence that D-brane charge takes values in the K-theory of the
spacetime manifold as before suspected in \moore.  Non-BPS states of some
brane configurations are classified according to the Type of superstring theory we are dealing. For
instance, in Type IIB theory, non-BPS states and their lower dimensional BPS-branes are classified by
the {\it complex} K-theory group K$(X)$
with $X$ the spacetime
manifold. In Type I theory non-BPS objects are classified by the {\it real} K-theory group
KO$(X)$. Non-BPS states in Type IIA (and their possible lifting to M-theory) are not well understood, however these
states are
classify by K$(X\times {\bf S}^1)$. Among other further natural generalizations,
Witten argued in
\witten\
that the correct description of non-BPS states in orbifold singularities is given in terms
of the {\it equivariant} K-theory (for a review of equivariant K-theory see \refs{\as,\segal,\greenless}).

In this paper we continue this analysis of configurations of pairs of
branes-antibranes in string theories
described by K-theory. In particular we give the first steps to study configurations
of D-brane pairs in orbifold singularities. We will see that equivariant K-theory
is the natural language to describe these configurations, just as it was argued  in
\witten. We find that many results, including the global construction of \witten, hold
also in the equivariant case with minor changes. However, we shall stress the differences
and explain some subtleties.

After
the previous discussion we apply this formalism to the specific case of
six-dimensional gauge theories on the world volume of Type IIB and I BPS-fivebranes (constructed from nine-brane
pairs) on a
general orbifold singularities. Although the construction is valid for general orbifolds, in order to
connect with some well known results of branes in orbifolds, we will restrict ourselves to the case of
branes on ALE singularities \refs{\dm,\johnson,\six,\bi}, leaving the
theories of D3 branes on ${\bf C}^3/\Gamma_G$ \dgm\ and D1 branes on ${\bf C}^4/\Gamma_G$
\mohri\ for
further investigation.
We have placed coincident fivebranes at at fixed point of the ALE singularity parametrized by the transverse
coordinates to the fivebrane.
Following Witten \witten, the fivebrane charge is now determined by the equivariant K-group of the compact space
of the transverse directions to the fivebranes or (in the non-compact case) by the corresponding
K-group with compact support. The transverse space is given by a ALE space which is
seen as the minimal resolution of the orbifold ${\bf C}^2/\Gamma_G$ with $\Gamma_G$ a discrete
subgroup of SU(2). We will make contact with the Kronheimer-Nakajima construction of Yang-Mills instantons
on ALE spaces \refs{\k,\kn}. We  show that all information concerning the description of
non-trivial RG fixed points of six-dimensional
theories in ALE singularities \refs{\seiberg,\six,\bi} in Type I and IIB theories, is contained in the
K-theoretic description and thus it admits a derivation in pure K-theoretic grounds.

The structure of this paper is as follows, in Section 2 we briefly review  Witten description of
non-BPS states in terms of K-theory. Section 3 is devoted to study non-BPS states in
orbifold singularities and their classification in terms of equivariant K-theory. In Section
4 we discuss in detail the problem of BPS-fivebranes in ALE singularities in Type I and IIB
superstring
theories. We find a nice relation with Kronheimer-Nakajima construction of instantons
and the description of nontrivial RG fixed points of theories in six dimensions
is derived from the mathematical
formalism of equivariant K-theory. In the process we discuss the origin of the index theorem
for ALE manifolds from the equivariant K-theory. In Section 5 we will provide the formula for the charge of
the brane in orbifold singularities derived from
K-theoretical considerations. Finally is Section 6 our concluding remarks are given.

\vskip 2truecm

\newsec{K-theory Description of Pairs D-brane-anti-D-brane}

In this section we briefly review the Witten's construction of D-brane pairs and their
relation to topological K-theory following \witten. Our aim is not provide
and extensive review, but only recall the relevant structure which will be needed in the
following sections. Throughout this paper we will follow the notation introduced by Witten in
\witten.


\subsec{Review of Witten's Construction}

\noindent
{\it K-Theory Structure}

In order to fix some notation let $X$ be the ten-dimensional spacetime manifold
and let $W$ be a $(p+1)$-dimensional submanifold of $X$.  Branes or antibranes  or both together can
be wrapped on $W$. When configurations of $N$ coincident branes or antibranes only
are wrapped on $W$, the world-volume spectra
on $W$ consists of a vector multiplet and scalars
in some representation of the gauge group.
These configurations can be described through Chan-Paton bundles which are U$(N)$ gauge bundles
$E$ over $W$ for Type II superstring theory and by SO$(N)$ or Sp$(N)$ bundles in
Type I theory. Gauge fields from the vector multiplet define a U$(N)$ gauge connection
for Type II theory (or SO$(N)$ or Sp$(N)$ gauge connection for Type I theory) on the
(corresponding)
Chan-Paton bundle. GSO projection cancels
the usual tachyonic degrees of freedom. Something similar occurs for the anti-brane sector.

The description of coincident $N_1$ $p$-branes and $N_2$ $p$-antibranes wrapped on
$W$ leads to the consideration
pairs of gauge bundles $(E,F)$ (over $W$) with their respective gauge connections
$A$ and $A'$. In the mixed configurations
GSO projection fails to cancel the tachyon. Thus the system is unstable and may flow toward
the annihilation of the brane-antibrane pairs with  RR charge for these brane configurations
being conserved in the process \refs{\sen,\senuno,\senew,\witten}.

On the open string
sector Chan-Paton factors are $2\times 2$ matrices constructed from the possible open strings
stretched among the different types of branes. Brane-brane and antibrane-antibrane sectors
correspond to the diagonal elements of this matrix. Off-diagonal elements correspond with
the Chan-Paton labels of an oriented open string starting at a brane and ending at an antibrane and
the other one to  be the open string with opposite orientation.

The physical mechanism of  brane-antibrane creation or annihilation without violation of
conservation of the total RR charge, leads to consider physically equivalent configurations
of $N_1$ branes and $N_2$ antibranes and the same configuration but with additional created
or annihilated brane-antibrane pairs.

The relevant mathematical structure describing the brane-antibrane pairs in general type
I and II superstring theories is as follows:

\item{i)-.} $G_1$ and $G_2$ gauge connections $A$ and $A'$ on the Chan-Paton bundles $E$
and $F$ over $W$, respectively. Bundles $E$ and $F$ corresponding to branes and antibranes
are topologically equivalent. The groups $G_1$ and $G_2$ are restricted to be unitary groups for
Type II theories and symplectic or orthogonal groups for Type I theories.

\item{ii)-.} Tachyon field $T$ can be seen as a section of the tensor product of bundles
$E \otimes
F^*$ and its conjugate $\bar{T}$ as a  section of $E^* \otimes F$ (where $*$ denotes the dual of the
corresponding bundle.)

\item{iii)-.} Brane-antibrane configurations are described by pairs of gauge bundles $(E,F)$.

\item{iv)-.} The physical mechanism of  brane-antibrane creation or
annihilation of a set of $m$ $9$-branes and $9$-antibranes is
described by the same U$(m)$ (for Type II theories) or SO$(m)$ (for type I theories)
gauge bundle $H$. This mechanism is described by the identification of pairs of gauge bundles
$(E,F)$ and $(E\oplus H, F \oplus H)$. Actually instead of pairs of gauge bundles one should
consider
classes of pairs of gauge bundles $[(E,F)]= [E] - [F]$ identified as above. Thus the brane-antibrane
pairs really
determine an element of the K-theory group K$(X)$ of gauge bundles over $X$
and the brane-antibrane creation
or annihilation of pairs is underlying the $K$-theory concept of {\it stable
equivalence} of bundles\foot{ For 9-branes, the embedded submanifold $W$ coincides
with $X$ and the thus brane charges take values in K-theory group of $X$.}.

\vskip .5truecm

Consistency conditions for 9-branes ($p=9$) in Type IIB superstring theory such as tadpole
cancellation implies the
equality of the ranks of the structure groups of the bundles $E$ and $F$. Thus
$rk(G_1)=rk(G_2)$. The `virtual dimension'
$D$ of an element $(E,F)$ is defined by $D= rk(G_1) - rk(G_2)$. Thus tadpole cancellation
leads to a description of the theory in terms of pairs of bundles with virtual
dimension vanishing, $D=0$. This is precisely the definition of reduced K-theory,
$\tilde{\rm K}(X)$. Thus consistency conditions implies to project the description to
reduced K-theory.

In Type I string theory $9-\overline{9}$ pairs are described by a class of pairs $(E,F)$ of
SO$(N_1)$
and SO$(N_2)$ gauge bundles over $X$. Creation-annihilation is now
described through the SO$(k)$ bundle $H$ over $X$. In Type I theories tadpole cancellation
condition is
$N_1-N_2 =32$. In this case equivalence class of pair bundles $(E,F)$ determines
an element in the {\it real} K-theory group KO$(X)$. Tadpole cancellation $N_1-N_2=32$,
newly turns out
into reduced real K-theory group $\tilde{\rm KO}(X)$.

Type IIA theory involves more subtle considerations worked out in \witten. It was argued by
Witten in \witten\ that configurations of brane-antibrane pairs are classified by the K-theory
group of spacetime with an additional circle space ${\bf S}^1 \times X$. K-theory
group for type IIA configurations is K$({\bf S}^1 \times X)$.

Finally, it is also shown in \witten\ that for both Type I and II theories,  the
consideration of
non-compact usual spacetime $X = {\bf R}^4 \times Q$ (where $Q$ some compact space) with
appropriate boundary conditions
at infinity leads to vacuum configurations with branes. In the language of K-theory
this means that a non-zero K-theory class $[(E,F)]$  is equivalent  to the
trivial class at infinity. Non-compact spacetimes $X$ require K-theory with compact support,
but with the boundary conditions at infinity the difference between ordinary and reduced K-theory
is irrelevant for the most of physical applications.

\break


\noindent
{\it Lower Dimensional Branes}

We now review Witten's global construction of lower dimensional $p$-branes, with $p < 9$ from
higher-dimensional branes. The flat case will be recovered when it is required
only one global coordinate system. The key argument is that it is possible to construct
$p$-branes as the bound state of $2^{n-1}$ pairs of $(p+2n)$-branes and antibranes
\refs{\senuno,\witten}.

Let $Z$ be a $(p+1)$-dimensional closed orientable submanifold of $Y$. This latter
is also a $(p+1+2n)$-dimensional orientable submanifold of spacetime manifold $X$ of the Type II
superstring theory. $Z$ is a codimension $2n$ submanifold of $Y$.
 $Y$ could possess a gauge
bundle ${\cal L}$ with a section vanishing along $Z$. Now we consider a system of
$N_1$ $(p+2n)$-branes and $N_2$ $(p+2n)$-antibranes wrapped on $Y$. Tachyon field $T$ of this
system transforms in
the bifundamental representation $(\fund_1, \antifund_2)$
of the group  ${\rm U}(N_1) \times {\rm U}(N_2)$. Tadpole cancellation condition $N_1=N_2=N$
implies that $T$ actually transforms in the adjoint representation $(\fund, \antifund)$ of
${\rm U}(N) \times {\rm U}(N)$. In vacuum tachyon field breaks the group
${\rm U}(N) \times {\rm U}(N)$ to the diagonal ${\rm U}(N)$. Tachyon is a section of the
bundle ${\cal L}$ and it is vanishing in a codimension $2n$ submanifold $Z$. Tachyon condensation
flows the system to a configuration with $N$ charges transforming in the diagonal
group ${\rm U}(N)$, representing the remaining $p$-branes.

The description of lower-dimensional
branes wrapped on $Z$ can be nicely described in terms of the pairs of brane-antibrane
wrapped in $Y$ as follows. The full formalism requires of the introduction of a U$(N)$ bundle
${\cal W}$ over $Z$. In order to extend the definition of this bundle from $Z$ to $Y$ several
obstructions can arise. For instance, if $W$ has trivial normal bundle ${\cal W}$ it can extends
over $Y$ getting an element of K$(Y)$ given by the class of pairs
$[({\cal L}\otimes{\cal W},{\cal W})]$.
However if ${\cal W}$ does not extend over $Y$ a more involved definition of the element of
K$(Y)$ has to be given using the standard tools of K-theory \refs{\atiyah,\bott,\karo,\spin}.

 Consider a tubular neighborhood $Z'$ of $Z$ and its closure $\overline{Z}$.
Bundles $({\cal L}\otimes{\cal W},{\cal W})$ over $Z$ can be easily pulled back
to $\overline{Z}$ determining so an element of K$(\overline{Z})$.
The corresponding
tachyon field $T: {\cal W} \to {\cal L}\otimes{\cal W}$ defines an isomorphism
when it is restricted to the boundary of $Z'$. Thus one can extend ${\cal W}$
over $Y'=Y-Z'$ via the map $T$ by declaring that this isomorphism is extendible over
$Y'$. If ${\cal W}$ does not extend over $Y'$ one can look for a gauge bundle $H$ over $Z$ in such a way that
${\cal W}\oplus H$ be trivial over $Z$ and trivial when pulled back to $\overline{Z}$.
Thus one can extend ${\cal W}\oplus H$ over
$Y$ (as the trivial bundle) and extend $ {\cal L}\otimes{\cal W}\oplus H$ by setting it equal to
${\cal W}\oplus H$ over $Y'$. The element of K$(X)$ is thus $({\cal L}\otimes{\cal W}\oplus H,
{\cal W}\oplus H)$ and the tachyon field is now $T\oplus1$.

\break

\noindent
{\it The Spinor Description}

Global construction outlined in the above subsection in general might not exist. However
it is possible to implement the global description in terms
spin bundles associated to the normal bundle to $Z$. Consider the normal bundle $N$ to
$Z$ in $X$. Assume that the codimension of $Z$ is $2n$. Normal bundle has structure group
SO$(2n)$. If $N$ is a spin bundle it has associated two spinor bundles $S_+$, $S_-$ of positive
and negative chirality spin representations of SO$(2n)$ respectively. If we suppose that the
spinor bundles near $Z$ extend over X, thus there are $2^{n-1}$ pairs of
$9-\overline{9}$
branes with gauge bundles $S_+$ and $S_-$ over $X$.

Tachyon field $T$ is a map given
by $T: S_- \to S_+$ and it is naturally
given by $T= \vec{\Gamma}\cdot \vec{x}$ where $\vec{x} = (x_1, \dots , x_{2n})$, where
$\vec{x}$ is an element of the tubular neighborhood $Z'$ of $Z$ in $X$. This tachyon map gives
a unitary isomorphism in the boundary of $Z'$. This isomorphism is determined  by the fact that
$\vec{\Gamma}\cdot \vec{x}$ is unitary if $\vec{x}$ is a unitary vector. One can use this
fact to extend this configuration over $X$ and the results of the above subsection
hold. Thus more generally one can look for a suitable bundle $H$ such that $S_-\oplus H$ is trivial
over $\overline{Z}$ and extends over $X$ with the final configuration given by
$(S_+\oplus H,S_-\oplus H)$ and the tachyon $T \oplus 1$.

\vskip 2truecm

\newsec{Equivariant K-theory Description of Branes in Orbifolds}

In the last section we gave the basic facts of the interplaying between D-branes configurations
and K-theory.  We consider now
brane-antibrane configurations in a generic orbifold and its description in terms of equivariant
K-theory. We will show that equivariant K-theory is the natural language to describe supersymmetric
and non-supersymmetric brane configurations in orbifolds.

\vskip 1truecm
\subsec{ Non-BPS States From  Brane-antibrane Pairs}

Before discussing BPS and
non-BPS branes in orbifolds, we briefly review the necessary requirements about
non-BPS states string theory on compact manifolds. We mainly follow the results
\refs{\sen,\senuno,\senew}. The basic idea is the construction of stable non-BPS
states through the construction of brane-antibrane  configurations. More precisely,
Sen has shown that wrapping branes and antibranes in cycles of some compact manifold in
some Type I and II superstring compactifications, world-volume theory has a tachyon
mode expansion which survives GSO projection. Tachyon field is an scalar which has
associated a negative potential energy
which cancel the tension between the brane-antibrane pair \refs{\sen,\senuno}.
 It was conjectured than unstabilities associated with the
tachyon mode flows the system to the annihilation of RR charge and
to an stable state admitting exact boundary conformal field theory description. To be more precise,
for instance, in Type I theory pairs of D1-brane- anti D1-brane have a tachyonic kink potential and this
system turns out to be stable and coincides with the SO(32) spinor state of the SO(32) heterotic theory
\senuno.

Pairs of D1-strings and anti-D1-strings  compactified on the circle
${\bf S}^1$ of radius $R$ have a tachyon field of the form $T(x)= \sum_{- \infty}^{+\infty}
T_n e^{i {n \over R}x}$. At the self-dual radius tachyon components
$T_{\pm}$ becomes massless and (in the $-1$ and zero picture) the vertex operators of the boundary CFT
are respectively given by

\eqn\tach{ V_{\pm}^{(-1)} = - e^{-\Phi_B} e^{\pm {i\over \sqrt{2}}X_B} \otimes \sigma_1}
and
\eqn\otrotach{ V_{\pm}^{(0)} = \mp i \psi_B  e^{\pm {i\over \sqrt{2}}X_B} \otimes \sigma_1,}
where $X_B$ is the bosonic field, $\Phi_B$ is the bosonized ghost and $\psi_B$ is the
world-sheet fermion. All of them taking its boundary value. $\sigma_1 = \pmatrix{0&1\cr
1&0\cr}$ is the Chan-Paton factor of the brane-antibrane pair open string sector.
Brane-antibrane configuration survives
$\Omega$ projection and it does exist in Type IIB theory \senuno.

\vskip 1truecm
\subsec{Branes-antibranes in Orbifolds}

Here we describe briefly the dynamics of pairs of branes-antibranes
in orbifold singularities. We present the case
of a pair of D-string-anti-D-string of the type IIB superstring theory
in a ${\bf Z}_2$-orbifold \senew. For definiteness take the pair with
world-volume along coordinates $(x^0,x^9)$ leaving transverse coordinates to be $(x^1,\dots,
x^8)$. ${\bf Z}_2$ group act only reversing the coordinates $(x^6,x^7,x^8,x^9)$ and leave
fixed the rest of the coordinates {\it i.e.} $(x^0, \dots, x^5)$. Thus action of ${\bf Z}_2$
behaves as an $A_1$ ALE singularity in the coordinates $(x^6,x^7,x^8,x^9)$. One can redefine
complex coordinates $z_1 =x^6+ix^7$ and $z_2=x^8 +ix^9$ and write down this singularity as
${\bf C}^2/{\bf Z}_2$. This is, of course, the most simple non-trivial example of the most simple class
of singularities of the {\bf A}-{\bf D}-{\bf E}-type.  In order to see how the brane pairs are behaved
at orbifold singularities we need to know how the orbifold projection is realized on the
Chan-Paton sector. As was shown in \senew, the Chan-Paton $I$ is even with respect the
action of ${\bf Z}_2$, while the Chan-Paton factor $\sigma_1$ is odd under ${\bf Z}_2$-orbifold
projection. This is essentially as transforms the tachyon field under the generator
$(-1)^F$ in \witten. Thus the tachyon is odd under the orbifold projection. We can
immediately generalize the ${\bf Z}_2$ group to $A_{N-1}$ singularities generated by the
group ${\bf Z}_N$ or for a generic group $\Gamma_G$ of the {\bf A}-{\bf D}-{\bf E} singularities.

We know from the basic theory of D-branes in orbifold singularities \refs{\dm,\johnson}
that the action of $\Gamma_G$ on the Chan-Paton factors is given by
$g(x(i)) = x(\gamma(g)(i))$
where $g$ is an element of $G$, $\gamma(g)$ of $\Gamma_G$ and $i$ is the index on the
Chan-Paton factors, which for the tachyon is $\sigma_1$. The tachyon field transforms under
$\Gamma_G$-orbifold projection as

\eqn\transtac{ g: T(x) \to \gamma(g) T(x') \gamma^{-1}(g).}
Thus similar than the vector potential $A_\mu(x)$, the tachyon $T$ transforms in the
adjoint representation of $\Gamma_G$ under orbifold projection.

\vskip 1truecm

\subsec{Equivariant  K-Theory Structure}

In this subsection we shall show in detail that the natural language to deal with
stable (non-)BPS branes in Type I and II superstring theory is, as suggested by Witten
\witten, the equivariant K-theory. Throughout the rest of the paper we limit ourselves
to work with $\Gamma_G$ an abelian and discrete subgroup of some {\bf A}-{\bf D}-{\bf E}
group.

To be the most self-contained possible we first give some basic definitions of the theory of
equivariant K-theory. For
details of proofs we encourage the reader to consult \refs{\as,\segal}.
Let $G$ be a Lie group, in general $G$ may be a topological group \foot{ The group $G$ does not be confused
with the generic ${\bf A}-{\bf D}-{\bf E}$ group.}. Let $X$ be a differentiable
manifold of finite dimension. $X$ is said to be $G$-manifold if there exist an
smooth action of $G$ on
$X$. A $G$-map between two $G$-manifolds is a smooth map which commutes with the action of the
group $G$.
Now consider a principal bundle $E$ over $X$ with canonical projection $\pi$. A $G$-principal
bundle $E_G$ over
the $G$-manifold $X$, is a $G$-map $\pi$ which carries fibres to fibres linearly and projects to the
action over $X$. A $G$-homomorphism $E_G \to F_G$ between two $G$-bundles over $X$ is a map which is both
a bundle homomorphism and a $G$-map. The $G$-isomorphism between two bundles $E_G \to F_G$ is a
$G$-homomorphism with inverse and the inverse is also a $G$-map.
Classes of pairs of $G$-bundles have a ring structure known as the Grothendieck ring
K$_G(X)$ or equivariant K-theory of $X$ \refs{\as,\segal,\greenless}.

When the group $G$ acts trivially on $X$, K$_G(X)$ factorizes just as the product
K$_G(X) = {\rm K}(X) \otimes R(G)$ where K$(X)$ is the ordinary K-theory ring and $R(G)$ is
the ring of irreducible representation spaces of $G$. Let $V_i$ be the $i-th$ component of the
irreducible vector space representation. For trivial actions $E_G$ can be written as
$E_G = \oplus_i Hom_G(\underline{E}_i,E) \otimes \underline{E}_i$ where $\underline{E}_i$ is the
trivial bundle $\underline{E}_i = X \times V_i$.

One of the properties of K$_G(X)$ we will use in this paper is the concept of space of
sections of a $G$-bundle or $G$-section space. Let ${\cal C}(E)$ be the space of sections of an
ordinary bundle $E$ over $X$, ${\cal C}(E)$ has the structure of a vector space and if $X$
is compact ${\cal C}(E)$ has in addition the structure of a Banach space. Now in the case
of we have a $G$-bundle $E_G$, in addition we have an induced action of $G$ on its space of
sections ${\cal C}(E_G)$. This action is continuous an it is given by $(g\cdot s)(x) = g \cdot s(g^{-1} \cdot x)$
with $g\in G$ and  $s \in {\cal C}(E_G)$.  To see that this action is continuous consider the
sequence of maps: $G \times {\cal C}(E_G) \times X \to G \times {\cal C}(E_G) \times X \to
G \times E_G \to E_G$ given by $(g,s,x) \mapsto (g,s,g^{-1}x) \mapsto (g,s(g^{-1}x)) \mapsto
(g\cdot s)(x)$. Each map is continuous and as composition of continuous mappings is also
continuous therefore this action is continuous. Action of $G$ over ${\cal C}(E_G)$ induces the group
homomorphism $J: G \to Aut({\cal C}(E_G))$.

Let $X/\Gamma_G$ be the ten-dimensional spacetime orbifold
and let $W/\Gamma_G$ be a $\Gamma_G$-submanifold (of dimension $p+1$) of $X/\Gamma_G$.
Let $j:W/\Gamma_G \hookrightarrow
X/\Gamma_G$ be an
embedding of $W/\Gamma_G$ into $X/\Gamma_G$ which preserves the action of $\Gamma_G$.
 Branes or antibranes can be now wrapped on $W/\Gamma_G$.
Thus the world-volume spectra on $W/\Gamma_G$ come from the dimensional reduction
of the corresponding string theory in ten dimensions with an additional orbifold projection.
The spectra
can thus be encoded in a {\it quiver diagram} as usual in brane theory in orbifolds
\refs{\dm,\johnson,\dgm,\mohri}.
The theory on the world-volume $W$ of $N$ branes (or antibranes) wrapped on $W$
is described through Chan-Paton bundles which are  given by
$\prod_\mu {\rm U}(Nn_\mu)$ bundles
$E$ over $W$, in the case of Type II superstring theory and by $\prod_{\mu}{\rm SO}(Nn_\mu)
\times \prod_{\mu}{\rm SU}(Nn_\mu)$ for the
Type I theory, here $n_\mu$ is the dimension of the $R_\mu$ regular representation of $\Gamma_G$.
Gauge fields from the vector multiplet defines a connection $A$ on the corresponding
Chan-Paton bundle. This connection satisfies, as usual, the orbifolding condition
$A(x)= \gamma(g)A(x') \gamma^{-1}(g)$, with $\gamma(g) \in \Gamma_G$.
Just as in the smooth case, GSO projection cancels
tachyonic degrees of freedom leaving only the quiver structure of the vector multiplet and
hypermultiplets.

When we consider both coincident branes and anti-branes wrapping on $W/\Gamma_G$ tachyon is still
preserved by GSO projection (as we have seen in the above subsection) and the vector multiplet is
also projected out as the smooth case. In other
words, in non-pathological cases, $\Gamma_G$-action commutes with GSO projection.
As we saw in the above subsection
tachyon field $T$ is required to satisfy the condition $T(x)=\gamma(g) T(x') \gamma^{-1}(g)$ with
$\gamma(g)$ being an element of $\Gamma_G$.

In terminology of {\it equivariant} K-theory, orbifold spacetime $X/\Gamma_G$ are seen as a
$\Gamma_G$-space {\it i.e.} spacetime $X$ with the action of $\Gamma_G$ over $X$.
Chan-Paton bundles are $\Gamma_G$-bundles $E_{\Gamma_G}$ over $X$ where the projection
map preserves the action of group $\Gamma_G$.  The tachyon
field $T$ can be seen as a $\Gamma_G$-map bundle $T: F_{\Gamma_G} \to
E_{\Gamma_G}$. Equivalently, tachyon can be seen now as a
$\Gamma_G$-section (up to a constant) of the $\Gamma_G$-bundle
$(E\otimes F^*)_{\Gamma_G}$. The way that tachyon field
transforms under $\Gamma_G$ given in \transtac, is nicely explained by a consequence of
$\Gamma_G$-section definition as follows.

Let ${\cal C}((E\otimes F^*)_{\Gamma_G})$ be the space of $\Gamma_G$-sections defined as
the space of sections of the $\Gamma_G$-bundle $(E\otimes F^*)_{\Gamma_G}$. The
group $\Gamma_G$ acts on ${\cal C}((E\otimes F^*)_{\Gamma_G})$ given by

\eqn\acttac{(\gamma(g) \cdot T) (x)
\equiv \gamma(g) \cdot T(\gamma^{-1}(g) \cdot x)= \gamma(g) \cdot T \cdot \gamma^{-1}(g) (x)}
 with $\gamma(g)$ and $T$ being elements of
$\Gamma_G$ and ${\cal C}((E^*\otimes F)_{\Gamma_G})$ respectively and $x\in X$. The action of
$\Gamma_G$ on sections ${\cal C}((E\otimes F^*)_{\Gamma_G})$ induces also a group homomorphism between
$\Gamma_G$ and $ Aut \big({\cal C}((E\otimes F^*)_{\Gamma_G}) \big)$.

When one is considering pairs of $9-\overline{9}$ branes on orbifold singularities similar
statements of the smooth case \witten\ are still valid. Creation and annihilation of
9-branes in orbifolds is described by a $\Gamma_G$-bundle $H_{\Gamma_G}$ over $X$.
So, still conservation of total charge, now including mirror images under $\Gamma_G$, leads
to make the identification
of pairs of bundles $(E_{\Gamma_G},E_{\Gamma_G})$ with $(E_{\Gamma_G} \oplus H_{\Gamma_G},
E_{\Gamma_G}\oplus H_{\Gamma_G})$, as two equivalent descriptions.
These conditions determine precisely an element of the $\Gamma_G$-equivariant K-group
${\rm K}_{\Gamma_G}(X)$ of spacetime $X$. Thus
brane-antibrane configurations in spacetime orbifold  $X/\Gamma_G$ can be described in terms
of equivariant K-theory methods. Thus D-brane charges on an orbifold singularity takes values
in the equivariant K-theory group of the spacetime $X$ as suggested in \witten.

In the specific case of Type IIB theories on orbifold singularities, tadpole cancellation
leads, just as the smooth case \witten, to the same number $N_1$ of 9-branes and $N_2$ of
$\overline{9}$-branes. The reason of this can be seen from the condition of equality of ranks of the gauge
groups

\eqn\tadpole{ \sum_\mu N_1 n_\mu = \sum_\mu N_2 n_\mu}
it is immediate to get $N_1=N_2=N$. Thus Chan-Paton bundles $E_{\Gamma_G}$ and
$F_{\Gamma_G}$ are $\prod_\mu {\rm U}(Nn_\mu)$ gauge bundles and tadpole cancellation
condition implies that virtual dimension is zero and it lies in equivariant reduced K-theory
group $\tilde{\rm K}_{\Gamma_G}(X)$.

In Type I string theory $9-\overline{9}$ pairs on orbifolds are described by a class of
pairs $(E_{\Gamma_G},F_{\Gamma_G})$ of
$\prod_{\mu}{\rm SO}(N_1n_\mu)$
and  $\prod_{\mu}{\rm SO}(N_2 n_\mu)$ gauge bundles over $X$. Creation-annihilation is now
described through the $\prod_{\mu}{\rm SO}(k n_\mu)$ bundle $H$ over $X$ for some $k$.
In Type I theories tadpole cancellation
condition is given by $\sum_\mu N_1n_\mu -  \sum_\mu N_1n_\mu= \sum_\mu 32n_\mu$. Thus condition
$N_1-N_2 =32$ holds again in this situation. Equivalence class of pairs of bundles
$(E_{\Gamma_G},F_{\Gamma_G})$ determines
an element in the {\it real} equivariant K-theory group KO$_{\Gamma_G}(X)$. Tadpole
cancellation $N_1-N_2=32$, newly turns out
into equivariant reduced real K-theory group $\tilde{\rm KO}_{\Gamma_G}(X)$.

Vacuum at infinity is $\Gamma_G$-invariant,
that means that it is reached by tachyon condensation preserving such a property.  The reason of this is
that vacuum space configuration of $T'$s is always written in an U$(N) \times {\rm U}(N)$ invariant
way \senuno, thus U$(N) \times {\rm U}(N)$ is broken down to the product
group $\prod_{\mu}{\rm U}(Nn_\mu) \times \prod_{\mu}{\rm U}(Nn_\mu)$.
The vacuum is reexpressed now in an invariant way in terms or the trace in the regular representation
of $\Gamma_G$. Equivalence to vacuum at infinity means that near infinity bundles
$E_{\Gamma_G}$ and $F_{\Gamma_G}$ are $\Gamma_G$-isomorphic. As the smooth case,
the vacuum might have some branes and therefore a non-zero class of K$_{cpt \ \Gamma_G}(X)$ where
our spacetime $X$ is a non-compact orbifold with the generic form
${\bf R}^k \times Q^{10-k}/\Gamma_G$. For Type I theory with
$\Gamma_G={\bf Z}_N$ an abelian and discrete subgroup of SU(2) and $Q={\bf C}^2$, tadpole cancellation is
still 32 for the 9-brane charge. For fivebranes, tadpole
constraints can be written in
terms of $\Gamma_G$-invariant geometric terms. For instance, in the unbroken SO(32) theory, anomaly cancellation
implies the existence of a certain number of fivebranes $F$, which is $F = 24 -N$ for $N$ even, and $F =
24 -N + {1 \over N}$ for $N$ odd \julie. In general there is not a choice of $F$ to cancel anomalies and the
anomaly inflow mechanism from the bulk should be extended to theories of branes in orbifolds \julie.

In physical applications descriptions of brane configurations in non-compact spacetime,
ordinary and reduced equivariant K-groups K$_{cpt \ \Gamma_G}(X)$ and
$\tilde{\rm K}_{cpt \ \Gamma_G}(X)$ are equivalent. And the reason of this is that the
$\Gamma_G$-invariance of the vacuum at infinity is described by a non-zero class of pairs of
$\Gamma_G$-bundles which is equivalent to the trivial class of pairs of bundles which are $\Gamma_G$-isomorphic
at infinity. This only is possible if
the pair of $\Gamma_G$-bundles have equal rank. Thus virtual dimension is zero and ordinary and
 equivariant reduced K-groups describe the same physical situation.

Type IIA theory involves more subtle considerations worked out in \witten. It was argued by
Witten in \witten\ that configurations of brane-antibrane pairs are classified by the K-theory
group of spacetime with an additional circle space ${\bf S}^1 \times X$.  Equivariant K-theory
of Type IIA theory not will be discussed in the paper, however we make only some few comments.
K-theory group for type IIA configurations is K$_{\Gamma_G}({\bf S}^1 \times X)$. If
$\Gamma_G ={\bf Z}_2$ and this group acts trivially on space $X$ and non-trivially on
${\bf S}^1$ then K$_{{\bf Z}_2}(X)$ is isomorphic to K$(X\times {\bf S}^1/{\bf Z}_2)$.
It is tantalizing to speculate that equivariant K-theory might be a relevant tool to generalize the
K-theoretical description of
non-BPS branes to $E_8\times E_8$ heterotic superstring theory.

\vskip 1truecm

\subsec{\it Lower Dimensional Branes on Orbifolds. Global Construction}

Up to here we have described $9-\overline{9}$ pairs of branes in generic orbifold singularities.
However to see that the equivariant K-theory classifies also lower dimensional branes than
9, we will
attempt to implement to orbifolds the Witten's construction \witten.

For branes of lower dimensions in orbifold singularities in type II superstring theory  the description
is  similar
to that given in \witten. However for completeness
we include here this discussion. The key argument here is that
it is possible to construct $|\Gamma_G|$
$p$-branes as the bound state of $2^{n-1} |\Gamma_G|$ pairs of $(p+2n)$-branes and
antibranes.

First we overview some notation. Let $Z_{\Gamma_G}$ be a $(p+1)$-dimensional (and codimension
$2n$) closed orientable
$\Gamma_G$-submanifold of $Y_{\Gamma_G}$. This latter is a $(p+1+2n)$-dimensional
$\Gamma_G$-submanifold of $X$. Let ${\cal L}_{\Gamma_G}$ be a $\Gamma_G$-bundle
over $Y_{\Gamma_G}$ with $\Gamma_G$-section vanishing along $Z_{\Gamma_G}$. One can consider
a set of
$N_1$ branes and $N_2$ antibranes wrapped on $Z_{\Gamma_G}$ such that tachyon field $T$ is a
$\Gamma_G$-section which transforms in the bifundamental representation
$(\fund_1, \antifund_2)$ of the
gauge group $\prod_{\mu} {\rm U}(N_1 n_\mu) \times \prod_{\mu} {\rm U}(N_2 n_\mu).$ Tadpole
cancellation condition implies that tachyon field transforms in the adjoint representation
$(\fund, \antifund)$ of the group $\prod_{\mu} {\rm U}(N n_\mu) \times \prod_{\mu}
{\rm U}(N n_\mu).$

Now introduce a $\Gamma_G$-bundle ${\cal W}_{\Gamma_G}$ over $Z_{\Gamma_G}$ with structure group
$\prod_{\mu} {\rm U}(N n_\mu)$. As in the smooth case, bundles ${\cal L}_{\Gamma_G}
\otimes {\cal W}_{\Gamma_G}$ and ${\cal W}_{\Gamma_G}$ determine an element of
the equivariant K-group of $Z_{\Gamma_G}$. If $Z_{\Gamma_G}$ has trivial normal bundle, then
the group K$_{\Gamma_G}(Z)$ admits an extension to $Y$ and we get  K$_{\Gamma_G}(Y)$.
If it does not occurs, then we use the $\Gamma_G$-invariance of the vacuum at infinity
$T: {\cal W}_{\Gamma_G} \to {\cal L}_{\Gamma_G}\otimes{\cal W}_{\Gamma_G}$  described by a
$\Gamma_G$-isomorphism of bundles on the boundary of a tubular neighborhood  $Z'_{\Gamma_G}$
of $Z_{\Gamma_G}$ in $Y_{\Gamma_G}$. Thus K$_{\Gamma_G}(Z)$ extends to K$_{\Gamma_G}(Y)$
by declaring that this $\Gamma_G$-isomorphism extends throughout $Y'_{\Gamma_G}= Y_{\Gamma_G}
- Z'_{\Gamma_G}$. If it does not occurs, then we apply the $\Gamma_G$-equivariant version
\refs{\as,\segal}\ of the
argument given in \witten.  One  extends rather the trivial bundle $ {\cal L}_{\Gamma_G}\otimes{\cal W}_{\Gamma_G}
\oplus H_{\Gamma_G}$ because it is $\Gamma_G$-isomorphic (with the isomorphism given by $T_G \oplus 1$)
to  ${\cal W}_{\Gamma_G}\oplus H_{\Gamma_G}$ over $Y'_{\Gamma_G}$.

\vskip 1truecm

\subsec{\it Examples}

\noindent
{\it Type I zero and -1 Brane}

We consider the zero brane of Type I superstring theory. This is described, according to subsection
$3.4$ by the group KO$_{\Gamma_G}({\bf S}^9)$ or the equivariant compact support group KO$_{cpt \ \Gamma_G}
({\bf R}^{9})$. The transverse space to the zero brane is the spacetime orbifold ${\bf R}^{9}/\Gamma_G$. Assume
that the group $\Gamma_G$ acts on the spatial coordinates only, ${\bf R}^{9}$ leaving fixed the
$x^9$ coordinate. Thus symbolically, $\gamma(g)\cdot(x^9,\vec{x}) = (x^9,\gamma(g)\vec{x} \gamma^{-1}(g))$,
where $\vec{x}=(x^1, \dots,x^8)$. $\vec{x}$ vector of the transverse space can be rewritten in terms of complex
coordinates $\vec{z}=(z_1,z_2,z_3,z_4)\in {\bf C}^4$ and to be more specific let $\Gamma_G$ be the cyclic group ${\bf Z}_N$.
Then choice an action of ${\bf Z}_N$ over ${\bf C}^4$ to be $(z_1,z_2,z_3,z_4) \to
\gamma(g)\vec{z} \gamma^{-1}(g) \equiv (e^{2\pi i/N} z_1, e^{-2\pi i/N} z_2,
e^{2\pi i/N} z_3, e^{-2\pi i/N} z_4).$

Thus actually we have the zero brane living at an orbifold singularity
${\bf C}^4/\Gamma_G$. An element of KO$_{\Gamma_G}({\bf R}^9)$ is given by a pair of trivial gauge bundles
$(E_{\Gamma_G},F_{\Gamma_G})$ over ${\bf R}^{10}$.  The tachyon field $T$ is a $\Gamma_G$-isomorphism
near infinity and it is given by $T(x) = \sum_{i=1}^9 \Gamma_i x^i$. Separation of
coordinates into $(x^9,\vec{x})$
breaks explicitly the symmetry SO(9) down to SO(8). Under this decomposition tachyon field reads

\eqn\tachyonic{T = \pmatrix{\vec{\Gamma} \cdot \vec{x}& x^9 \cr
-x^9&\vec{\Gamma}^T \cdot \vec{x}\cr}}
where $\vec{\Gamma}$ are the SO(8) gamma matrices satisfying
$\vec{\Gamma}:S^-_{\Gamma_G} \to S^+_{\Gamma_G}$, $\vec{\Gamma}^T: S^+_{\Gamma_G} \to S^-_{\Gamma_G}.$
$\vec{\Gamma}^T$ is
transpose to $\vec{\Gamma}$ and $S^\pm_{\Gamma_G}$ are the
positive and negative SO(8) chirality $\Gamma_G$-spinor bundles over the unidimensional space.
Each $S^\pm_{\Gamma_G}$
bundle can be trivially extended over ${\bf R}^{9}$.
The system splits in two components:
$T_1= \vec{\Gamma_G} \cdot \vec{x}$ and $T_2= \vec{\Gamma_G}^T \cdot \vec{x}$. Recall that $\vec{x}$
are the coordinates of the orbifold and the action of $\Gamma_G$ on $\vec{x}$ will depends on the
explicit group $\Gamma_G$. $S^\pm_{\Gamma_G}$ are $\Gamma_G$-bundles and the SO(8) Dirac matrices $\vec{\Gamma}$ can be
seen as $\Gamma_G$-sections of the  $\Gamma_G$-bundle $(S_-\otimes S_+^*)_{\Gamma_G}$ and according to
the transformation
of $\Gamma_G$-sections, they
transform as $\gamma(g) \vec{\Gamma}_A \gamma^{-1}(g)$ with $\vec{\Gamma}_1= \vec{\Gamma}$ and $\vec{\Gamma}_2
=\vec{\Gamma_G}^T$. Under the combination $T_A= \vec{\Gamma_G}_A \cdot \vec{x}$, the transformation of $\vec{x}$ and
$\vec{\Gamma_G}$ implies that the tachyon field transforms correctly as given from Eq. (3.4)

\eqn\zero{ T_A \to \gamma(g) T_A \gamma^{-1}(g), \ \ \ \  A=1,2.}

Similar to the smooth case $T_1$ describes a set of $|\Gamma_G|$ D1-branes located at an
orbifold singularity at the origin of $\vec{x}$ and  $T_2$ describes the other set of $|\Gamma_G|$
D1-antibranes located at the origin of the orbifold singularity.  This orbifold is, as we have seen,
an orbifold of fourfolds given by ${\bf C}^4/\Gamma_G$.

On the other hand the study of the -1 brane in an orbifold singularity is very similar to the analysis for the zero
brane. In this case the system is classified by K$_{\Gamma_G}({\bf R}^{10})$ with compact support.
Now, there are two $\Gamma_G$-spinor bundles $S^{\pm}_{\Gamma_G}$ coming from the real spinor
representations of SO(10). Action of $\Gamma_G$ on $\vec{x}=(x^1,\dots,x^{10})$ determines a singularity
of the type ${\bf C}^5/\Gamma_G$ and thus there is only one tachyon field $T= \vec{\Gamma}\cdot \vec{x}$.

\vskip 1truecm

\subsec{\it Equivariant Bott Periodicity}

Equivariant Bott periodicity for KO$_{\Gamma_G}$-theory with compact support is given by

\eqn\boot{ {\rm KO}_{\Gamma_G}({\bf R}^n) = {\rm KO}_{\Gamma_G}({\bf R}^{n+8}).}
This formula tell us that $|\Gamma_G|$ 1-branes and $|\Gamma_G|$ 7-branes or $|\Gamma_G|$ 0-branes and
$|\Gamma_G|$ 8-branes in the orbifold singularity, are related. In order to see that consider a $(n-1)$-brane
wrapped on ${\bf R}^n$ and let $j:{\bf R}^{n} \hookrightarrow {\bf R}^{n+8}$ be an embedding. $n-1$ brane
is living in an orbifold singularity of the type ${\bf R}^8/\Gamma_G$, being the coordinates
$\vec{x}$ of ${\bf R}^8$
given by the last eight coordinates of ${\bf R}^{n+8}$. Let $(E^{(0)}_{\Gamma_G},F^{(0)}_{\Gamma_G})$ be an
element of KO$_{\Gamma_G}({\bf R}^n)$. Tachyon field is a $\Gamma_G$-map of bundles $T_0:F^{(0)}_{\Gamma_G}
\to E^{(0)}_{\Gamma_G}$. The embedding  $j$ has an associated normal bundle with structure group
SO(8). This have two $\Gamma_G$-spinor bundles $S^{\pm}_{\Gamma_G}$ of positive and negative chiralities
associated with the normal bundle. Using the methods of subsection 3.4 we can extend the above pair of bundles
over ${\bf R}^8/\Gamma_G$ and thus to get a class of pairs of bundles of
KO$_{\Gamma_G}({\bf R}^{n+8})$ given by $(E^{(0)}_{\Gamma_G}\otimes (S^+_{\Gamma_G}
\otimes S^-_{\Gamma_G}),F^{(0)}_{\Gamma_G}\otimes (S^+_{\Gamma_G} \otimes S^-_{\Gamma_G}))$ over
${\bf R}^{n+8}$. The associated tachyon field is given by

\eqn\tachyonic{T = \pmatrix{\vec{\Gamma} \cdot \vec{x}& T_0 \cr
-T_0&\vec{\Gamma}^T \cdot \vec{x}\cr}.}
As the above subsection, the system splits in two tachyonic components transforming under the regular representation of
$\Gamma_G$ and the full tachyonic field will transform under the 2-dimensional fundamental
representation of $\Gamma_G$. Thus
we have found that Bott periodicity relating branes in string theory is still valid when we sit them on a
generic orbifold singularity relating now not only these branes but also their $|\Gamma_G|$ mirror images
under $\Gamma_G$.

Bott periodicity in its equivariant version for complex K-theory is given by the natural
isomorphism \refs{\as,\segal,\sma}

\eqn\eqboyy{ {\rm K}_G^{-q}(X) \cong  {\rm K}_G^{-q-2}(X)}
where K$_G^{-q}(X)$ is defined by K$_G(\Sigma^qX)$ with $\Sigma^qX$ the $q$-th reduced
suspension of $X$. Thus equivariant Bott periodicity for complex K-theory has, as the
ordinary Bott periodicity, periodicity equal to 2. Thus for Type IIB superstring theories there are
lower dimensional
BPS-branes each two dimensions and Bott periodicity identifies the branes with dimension $q$ with
branes with dimension $q+2$. With $G= \Gamma_G$ essentially the same thing occurs when Type IIB
branes are placed in orbifold singularities of the type ${\bf C}^k/\Gamma_G$ with $k=2,3,4$.
As we have seen, the interpretation of
equivariant Bott periodicity theorem is the identification of sets of $|\Gamma_G|$ $q$-branes with
$|\Gamma_G|$ $(q+2)$-branes with $|\Gamma_G|$ the number of mirror images in the orbifold.
Later at the end of the Section 4 we will come back to this point.

\vskip 2truecm

\newsec{Fivebranes, Small Instantons and Theories in Six Dimensions}

As we have reviewed in Type I and IIB is possible to construct some stable D-branes as a bound
state of certain number of $9-\overline{9}$ pairs.
In this section we study the description of stable supersymmetric Type I and IIB fivebranes in a
background of nine-brane pairs as worked out in \witten. Now we sit the fivebrane in an ALE
orbifold singularity ${\bf C}^2/\Gamma_G$ in the transverse space of the fivebrane. We have choice
an ALE singularity in order to
connect with some previous results well known from the literature but in principle its generalization to other
orbifold singularities abelian or non-abelian is immediate. Our main claim is that the description of
fivebrane from $9-\overline{9}$ brane pairs in orbifold singulatities are classified by equivariant
K-theory group K$_G(X)$ with $G=\Gamma_G$. We present evidence of this by deriving the relevant
information to describe non-trivial RG fixed points on the fivebrane world-volume theory, through the
equivariant K-theory formalism.

\vskip 1truecm

\subsec{Fivebrane From Nine-brane Pairs}

As was shown in \witten, fivebranes can be interpreted in terms of
nine-brane pairs. In Type I theory, instantons are precisely constructed from bound
states of fivebrane and ninebranes \refs{\small,\douglas}. We consider four pairs of
$9-\overline{9}$ branes
and one fivebrane in Type I theory. This configuration describes a small instanton of charge one
and gauge group SU(2)=Sp(1). Instanton charge takes values in the group K$({\bf S}^4)$ or
K$({\bf R}^4)$ with compact support with  ${\bf R}^4$ parametrizing the transverse directions to the
fivebrane. This system is equivalently described by the global description of $Z={\bf S}^4$. It is easy to
see that the normal bundle $N$ to ${\bf S}^4$ in ${\bf R}^{10}$ is a bundle with structure group
SO$(2n)$ with $n=3$. Thus following the global constructions reviewed in Section 2, the fivebrane
can be constructed from $2^{n-1}$ pairs of 9-branes {\it i.e.} four pairs of 9-branes. Normal bundle
is a spin bundle and there is a pair of spinor bundles $S_\pm$ associated to the positive and
negative chiralities of SO(6). Tachyon field is given by $T = \vec{\Gamma}\cdot \vec{x}$
with $\vec{x}=(x_1, \dots, x_6)$ are a point of the fivebrane world-volume.
Tachyon field breaks the group SO$(4) \times {\rm SO}(4)$ to the diagonal SO(4), which is
isomorphic (locally) to SU$(2)\times {\rm SU}(2),$ this group is broken by the instantons to
SU(2).

One can attempt an immediate generalization of the last paragraph. Consider a set of $k$ fivebranes
constructed from $N=2^{n-1}k$ pairs of $9-\overline{9}$ branes. Following \small, one can
construct easily the configuration of $k$ instantons with gauge group SO$(N)$ with $N=4k$. Later
in
subsection $4.5$ we will discuss with detail the K-theoretic description of the small instanton
in an ALE singularity.

\vskip 1truecm

\subsec{ Type IIB Instanton in an ALE Orbifold Singularity}

We now consider fivebranes located at general {\bf A}-{\bf D}-{\bf E} type orbifold singularities of
Type IIB theory. Orbifold singularity we deal is of the form ${\bf R}^4/\Gamma_G$ where
$\Gamma_G$ is a discrete and abelian subgroup of SU$(2)$. Consider a configuration consisting of
$k$ fivebranes whose world-volume is parametrized by the coordinates $(x^0,\dots ,x^5)$ located
at a fixed point of the orbifold singularity
${\bf C}^2/\Gamma_G$, parametrizing the normal directions
to the fivebrane $(x^6,x^7,x^8,x^9)$. Fivebrane charge takes values in the equivariant
group K$_{\Gamma_G}({\bf S}^4)$ or equivalently in
K$_{\Gamma_G}({\bf R}^4)$ with compact support.
Moreover, the relevant group to
compute it is rather the K-theory group K$(Z_{\zeta})$ of the  ALE space $Z_\zeta$, with
$Z_\zeta$ being isomorphic to the minimal resolution of ${\bf R}^4/\Gamma_G$ \k.

\vskip 1truecm

\subsec{Relation to Kronheimer-Nakajima Construction}

In Type IIB theories we have seen that an element of the equivariant K-theory group
is given by a class of pairs of $\Gamma_G$-gauge bundles $(E_{\Gamma_G},F_{\Gamma_G})$
both are $G'$-bundles over $Z_\zeta$, with
$G'= \prod_{\mu=0}^r {\rm U}(N n_\mu)/{\rm U}(1)$ where $r=rank(G)$ and $G$ is some {\bf A}-{\bf D}-{\bf E}
group: $A_r$, $D_r$, $E_6$, $E_7$ and $E_8$. Group $\Gamma_G$
has the regular irreducible representation $R$. With the action of
$\prod_{\mu=0}^r {\rm U}(N n_\mu)$ over $R$, one can construct the associated
$\Gamma_G$-vector bundle
given by $R_i - E_{\Gamma_G} \times_{G'} R \to Z_\zeta$. Similarly for $F_{\Gamma_G}$ we can construct a
$\Gamma_G$-vector bundle $R_i - F_{\Gamma_G} \times_{G'} R \to Z_\zeta$.
One can immediately see that vector bundle ${\cal E}_{\Gamma_G}=E_{\Gamma_G} \times_{G'} R$ it is very well
known from the literature \kn\ and it is known as the {\it tautological bundle}. Thus we have
found that equivariant K-group determining the charge of $k$ fivebranes in an ALE
orbifold singularity is equivalently described as a class of pairs of tautological
bundles arising in Kronheimer-Nakajima construction of instantons \kn\ {\it i.e.}
K$_{\Gamma_G}(Z_\zeta) = ({\cal E}_{\Gamma_G},{\cal F}_{\Gamma_G})$. These bundles are
$\Gamma_G$-modules given by ${\cal E}_{\Gamma_G} = \oplus_{\mu =0}^r {\cal E}_{\Gamma_G \ \mu}
\otimes \underline{R}_\mu$ where $R_\mu =(R_0, \dots, R_r)$ are the irreducible regular
representations of $\Gamma_G$, ${\cal E}_{\Gamma_G \ \mu} = E_{\Gamma_G} \times_{G'} R_\mu$ and
$\underline{R}_\mu$ is the trivial bundle $\underline{R}_\mu= Z_\zeta \times R_\mu$ over $Z_\zeta$
with fiber $R_\mu$.

In order to extract physical information it is needed to describe the K-group K$_{\Gamma_G}
(Z_\zeta)$ in terms of more familiar grounds.
K-theoretical description can be usually translated in terms of cohomological language
through the famous index theorem of elliptic operators \ast. To do this Atiyah and Singer used
an alternative definition of equivariant K-theory in terms of complexes of $G$-bundles
over $G$-spaces \refs{\as,\segal}.  To
proceed further we need a tubular neighborhood of $Z_\zeta$ and the Thom isomorphism.  Tubular
neighborhood $Z'_\zeta$ of $Z_\zeta$ in ${\bf R}^{10}$ is familiar for us from the above sections so
 we will focus on
the Thom isomorphism.
One can construct the Thom isomorphism as
follows: Start from the cotangent bundle $T^*Z_\zeta$, to $Z_\zeta$ with canonical projection
$\pi$. The embedding of $Z_\zeta$ into
regular representation vector space $R$ induces an embedding of $T^*Z_\zeta$ into $T^*R$ and so the Thom isomorphism
is given by
$\psi: {\rm K}_{\Gamma_G}(T^*Z_\zeta) \to {\rm K}_{\Gamma_G}(\pi^*(Z'_\zeta \otimes_{\bf R} {\bf C}))$.
Also there is an embedding
$i:\pi^*(Z'_\zeta \otimes_{\bf R} {\bf C}) \hookrightarrow T^*R$ and its corresponding inducing map
$i_*: {\rm K}_{\Gamma_G}(\pi^*(Z'_\zeta \otimes_{\bf R} {\bf C})) \to {\rm K}_{\Gamma_G}(T^*R)$.
Thus one can construct the Gysin
map $i_!: {\rm K}_{\Gamma_G}(T^*Z_\zeta) \to {\rm K}_{\Gamma_G}(T^*R)$ as $i_!=i_* \circ \psi$. The
inclusion of a point
$P$ into $R$ induces similarly the map $j_!: {\rm K}_{\Gamma_G}(T^*P) \to {\rm K}_{\Gamma_G}(T^*R)$.
Thus $\Gamma_G$-equivariant topological index ${ind_t}:  {\rm K}_{\Gamma_G}(T^*Z_\zeta)
\to R(\Gamma_G)$ is defined as $ind_t=((j_!)^{-1} \circ i_!)([{\cal O}_{\Gamma_G}])$ where
${\cal O}_{\Gamma_G}$ is a
$\Gamma_G$-vector bundle over $T^*Z_\zeta$. This  is precisely the well known Atiyah-Singer
$\Gamma_G$-index formula \ast.
Equivariant Chern character $ch_{\Gamma_G}: K_{\Gamma_G}(Z_\zeta) \to
H_{\Gamma_G}(Z_\zeta;{\bf Q})$ is an useful tool to translate the index formula
in cohomological language \ast. Coincidence between topological and analytical index leads to the
Atiyah-Singer theorem for elliptic operators $D$ \ast
\eqn\index{ Ind (D) = - \int_{Z_\zeta} ch_{\Gamma_G}({\cal E}_{\Gamma_G}) \hat{A}(Z).}

As $Z_\zeta$ is a ALE manifold with non-trivial boundary ${\bf S}^3/\Gamma_G$, the $\Gamma_G$-index
theorem has
to be modified to $\Gamma_G$-index for manifolds with boundary \refs{\aps,\pope}.
For twisted Dirac operators
$D^{\pm}: {\cal C}(S^\pm \otimes {\cal E}_{\Gamma_G}) \to  {\cal C}(S^\mp \otimes {\cal E}_{\Gamma_G})$
the above equation has to be
corrected by a boundary term given by the $\eta$ invariant \kn

\eqn\index{ Ind (D) = - \int_{Z_\zeta} ch_{\Gamma_G}({\cal E}_{\Gamma_G}) \hat{A}(Z) + {1 \over |\Gamma_G|} \sum_{\gamma \not= 1}
{\chi_W(\gamma) \over 2 - \chi_Q(\gamma)}.}
This formula determines the dimension of the moduli space of instantons in ALE gravitational spaces \kn.
In cohomological terms the use of characteristic classes gives relevant information about the non-trivial
submanifolds of the ALE space. For instance the first Chern classes $c_1({\cal E}_{\Gamma_G \mu})$
of ${\cal E}_{\Gamma_G \mu}$ with $\mu =1,\dots,r$ form a basis of $H^2(Z_\zeta)$ \kn. Poincar\'e
duality determines $r$ non-trivial homology cycles $\Sigma_\mu=(\Sigma_1, \dots,\Sigma_r)$. The representation $R_\mu$
is strongly associated to the monodromy of ${\cal E}_{\Gamma_G \mu}$ at the end of $Z_\zeta$, $Z_{\infty}$.
In other words $\pi_1(Z_{\infty})=\Gamma_G$ and consequently there are non-trivial Wilson loops at
infinity providing a representation of $\Gamma_G$ into the gauge group.

Kronheimer-Nakajima construction of instantons on ALE spaces \kn, requires the construction of
$\Gamma_G$-vector bundle ${\cal J}_{\Gamma_G}$ over $Z_\zeta$ with anti-self-dual connection,
 from some ADHM data. This bundle is of course a generalization of our friend  ${\cal E}_{\Gamma_G \mu}$.
Among these data there are two
complex vector spaces $V$ and $W$ which are $\Gamma_G$-modules and can be written as $V=\oplus_{\mu=0}^r V_\mu
\otimes R_\mu$ and $W=\oplus_{\mu=0}^r W_\mu \otimes R_\mu$, of dimensions
$k=dim(V)=\sum_{\mu}n_\mu v_\mu$ and $N=dim(W)=\sum_{\mu}n_\mu w_\mu$ respectively. The mentioned bundle is
given by ${\cal J}_{\Gamma_G} = Ker(D^{\dag}_{\Gamma_G})= Coker(D_{\Gamma_G})$ where
$D^{\dag}_{\Gamma_G}: {\cal V} \oplus
{\cal W} \to {\cal U}$ where ${\cal V}= (\underline{Q}\otimes \underline{V} \otimes E)_{\Gamma_G}$,
${\cal W}= (\underline{W} \otimes E)_{\Gamma_G}$ and
${\cal U}= S^+\otimes (\underline{V} \otimes E)_{\Gamma_G}$.

The vector $\vec{w}=(w_0,\dots,w_r)$
contains information about the monodromy representation coming from the limiting flat connection on the
end of $Z_\zeta$. In this limit the bundle ${\cal J}$ is given by ${\cal J}_{\infty}= \oplus_{\mu=0}^r w_\mu
R_\mu$.

The first Chern class of ${\cal J}_{\Gamma_G}$ is given by
$c_1({\cal J}_{\Gamma_G}) = c_1({\cal V}) + c_1({\cal W}) -c_1({\cal U})=\sum_{\mu=0}^r u_\mu c_1({\cal E}_{\Gamma_G \mu})$
with $u_\mu = w_\mu - \sum_{\nu}\tilde{C}_{\mu \nu}v_\nu$ where $\tilde{C}_{\mu \nu}$ is the extended Cartan
matrix $\tilde{C}_{\mu \nu}= 2 \delta_{\mu\nu} - a_{\mu\nu}$ and  $a_{\mu\nu}$ is  the matrix constructed
from the extended Dynkin diagram of the corresponding {\bf A}-{\bf D}-{\bf E} group. $a_{\mu\nu}$ enters in the
representation formula $R_Q \otimes R_\mu = \oplus_{\mu=0}^r a_{\mu\nu}R_\nu$ with $R_Q$ the fundamental
two-dimensional representation of $\Gamma_G$. It is convenient rewrite the resulting formula from first Chern class
for future reference as

\eqn\chernfirst{ \tilde{C}_{\mu \nu} v_\nu = w_\mu -u_\mu.}

While the second Chern class  has two
contributions: the ordinary contribution due to the non-trivial topology of $Z_\zeta$ and
that of the asymptotic behavior. Thus second Chern class is given by
$c_2({\cal J}_{\Gamma_G}) = \sum_\mu u_\mu c_2({\cal E}_{\Gamma_G \ \mu}) + {k \over |\Gamma_G|}.$ The integrals
of $c_2({\cal J}_{\Gamma_G})$ determines
the instanton number $I=k({\cal J}_{\Gamma_G})= \sum_{\mu} u_\mu k({\cal E}_{\Gamma_G \ \mu}) + (dim V)/|\Gamma_G|.$

Finally, the moduli space ${\cal M}_{k,N}$ of $k$ U$(N)$ instantons is the space of anti-self-dual Yang-Mills instantons
on an ALE space. Construction of this moduli space and uniqueness was proved using the ADHM data by
Kronheimer and Nakajima \kn. The dimension of ${\cal M}_{k,N}$ can be determined by the Atiyah-Patodi-Singer
for manifolds with boundary \refs{\aps,\pope}. The result is $dim ({\cal M}_{k,N}) ={1\over 2}v_\mu(w_\mu +u_\mu)$.

\vskip 1truecm
\subsec{Type IIB Six-dimensional Gauge Theories From Fivebranes in ALE Orbifolds }

Six-dimensional theories on the world-volume of a type IIB $k$ coincident fivebranes in a general
{\bf A}-{\bf D}-{\bf E}
ALE singularity  and in a background of $N$ 9-branes,
have many interesting
properties \refs{\six,\bi}. One of the most interesting is the existence of non-trivial RG fixed points.
Some results are known but it remains to explore other descriptions of these points using new techniques.
Important dynamical information
is extracted from these points, which are described by a supersymmetric gauge theories with
eight supercharges and gauge group $\prod_{\mu=0}^r {\rm U}(kn_\mu)$. The theory without 9-branes has
${\cal N} =(1,0)$ supersymmetry
and their matter multiplets transforms as ${1\over 2} a_{\mu \nu}(\fund_\mu,\antifund_\nu)$. In addition there
are $r$ ${\cal N} =(1,0)$ hypermultiplets and $r$ ${\cal N} =(1,0)$ tensor multiplets. These come from the
six-dimensional reduction of the ten-dimensional two-form and four-form potentials on the $r$ two-dimensional
cycles $\Sigma_{\mu}= (\Sigma_1,\dots ,\Sigma_r)$ of $Z_\zeta$.

The Higgs branch
of this theory is shown to be isomorphic to the hyper-K\"ahler quotient construction of Yang-Mills instantons
in ALE gravitational instantons \refs{\dm,\six,\bi}. First Chern class is realized by the integral of NS-NS $B$
in the non-trivial cycles $\Sigma_\mu$ as $\int_{\Sigma_{\mu}} B = n_\mu /|\Gamma_G|$ with $(\mu=1,\dots,r)$.
In the first reference of \bi, it was shown that for Type I and II theories, worldsheet tadpole cancellation condition
for instanton configurations in ALE spaces implies the cancellation of gauge anomalies of the six-dimensional theory on the
world-volume of the fivebranes. Tadpole cancellation condition fivebranes in ALE spaces is shown to be
contained in the topological and group information of the ALE singularity through the formula (4.3).

The six-dimensional
theory on the fivebranes world-volume is anomaly free requiring exactly the $r$ hypermultiplets and the $r$
tensor multiplets to
cancel anomalies. The anomaly coming from the $r$ U$(1)$ factors of $\prod_{\mu=0}^r {\rm U}(kn_\mu)$ are
canceled by the $r$ mentioned hypermultiplets. The other contribution to the anomaly is canceled by the coupling of
the $r$ tensor multiplets to the gauge fields. Thus a sensible description of these theories requires
the full topological items coming from
Kronheimer-Nakajima construction.
From the point of view of equivariant K-theory, fivebranes configurations are completely
classified by the group K$_{\Gamma_G}(Z_\zeta)$ and as we have seen in this Section, an element of this
group can be expressed in terms of a pair of tautological bundles providing thus a connection to
Kronheimer-Nakajima construction. Thus all relevant topological information to describe anomaly
free gauge theories in six-dimensions is contained as we have shown,
in equivariant K-theory group K$_{\Gamma_G}(X)$ as suggested by Witten in \witten.

\vskip 1truecm
\subsec{Type I Small Instantons and Six-dimensional Gauge Theories}

Now we consider type I theory of  $k$ coincident fivebranes located at general ALE singularity
${\bf R}^4/\Gamma_G$ and in a background of $N$ pairs of $9-\overline{9}$ branes. These configurations
describe $k$ SO(N) small instantons on the ALE space.
The description of these configurations
is in much as the above subsection. We only will comments the differences and make some remarks. In
these type of theories there is two possible cases: the case with vector structure and the case
without vector structure. We submit the reader to \bi\ for details. Here we only consider the case with
vector structure.

In Type I theories we have seen that an element of the equivariant KO-theory group
is given by a class of pairs  of real $\Gamma_G$-gauge bundles $(E_{\Gamma_G},F_{\Gamma_G})$
both are $G'$-bundles over $Z_\zeta$, with gauge group
$G'= \prod_{\mu\in {\cal R}} {\rm Sp}(v_\mu) \times
\prod_{\mu\in {\cal P}} {\rm SO}(v_\mu) \times
\prod_{\mu\in {\cal C}} {\rm U}(v_\mu)$ were ${\cal R}$ denotes real, ${\cal P}$
pseudoreal, ${\cal C}$ complex and $\overline{\cal C}$ complex conjugate, representations of $\Gamma_G$.
With the action of
$G'$ over $R$, one can construct the associated vector bundle
given by $R_i - E_{\Gamma_G} \times_{G'} R \to Z_\zeta$ and similar for $F_{\Gamma_G}$.
Once again the vector bundle ${\cal E}_{\Gamma_G}=E_{\Gamma_G} \times_{G'} R$  is the
 tautological bundle. Thus the equivariant K-group determining the charge of $k$ fivebranes in an ALE
orbifold singularity is equivalently described as a class of pairs of tautological
bundles from the Kronheimer-Nakajima construction of instantons \kn.

KO-theoretical description can be translated in cohomological terms. Thus the basic formulas
are still valid. First Chern class leads to $\tilde{C}_{\mu \nu}$ $V_\nu = w_\mu -D_\mu$ where $V_\mu = {\rm dim}
(\fund_\mu)$ {\it i.e.} $V_\mu = 2 v_\mu$ for Sp$(v_\mu)$ and $V_\mu = v_\mu$ for SO$(v_\mu)$ and U$(v_\mu)$.
Instantons on ALE space are characterized by the instanton number $k$ which come from the Chern second class
and by the vector $\vec{w}=(w_0,\dots,w_r)$ which characterize the Wilson loops at the end of $Z_\zeta$ and
satisfying the condition $\sum_\mu n_\mu w_\mu =32$. The theory also possesses matter as hypermultiplets and
tensor multiplets which are associated by phase transitions of the six-dimensional gauge theory \bi.

Equivariant $\Gamma_G$ K-theory groups provide once again all topological and group theoretical
information to describe the Higgs and Coulomb branches of type I six-dimensional theory associated
to small instantons. Actually this is not surprising,
but what is interesting is that all relevant information is encoded from the beginning in the equivariant
KO-theory groups. Thus we claim that K-theoretical tools
enter deeply in the problem and may
provide of a very useful tool to study and classify non-trivial RG fixed points of six-dimensional theories.

\vskip 1truecm

\subsec{AdS/CFT Correspondence and  Orbifold Singularities}

In Type IIB string theory D-brane theories on orbifolds also can be realized in the
context of the AdS/CFT correspondence \ads. It is well known the equivalence of
superconformal field theories with various degrees of supersymmetry on the D-brane
world-volume is equivalently described by Type IIB D$(p-2)$-branes in orbifolds
$AdS_p \times S^{10-p}/\Gamma_G$. The field theory on the world-volume of a D brane probes
on singularities can be determined using open string techniques introduced in
\refs{\dm,\dgm}. In particular for $p=5$ one have a theory of Type IIB D3-branes with
supersymmetry on the world-volume ${\cal N}=2,1,0$ corresponding to $\Gamma_G$ being a
subgroup of SU(2), SU(3), SU(4) respectively \refs{\ks,\vafa}. Admissible values of $p$ are
given by the practice to be $p=3,5,7$.

We focus for the moment on the case $p=5$, this corresponds to $AdS_5 \times
S^5/\Gamma_G$. When $\Gamma_G$ is a subgroup of SU(2) of the $A_{N-1}$ type {\it i.e.}
$\Gamma_G ={\bf Z}_N$ then, the field theory on the world-volume of $n_1$ D3 branes
corresponds to a superconformal field theory in four dimensions with ${\cal N}=2$
supersymmetry and gauge group U$(n_1)^{|\Gamma_G|}$. For $\Gamma_G$ a subgroup of SU(3)
the supersymmetry on the world-volume theory of D3-branes is ${\cal N}=1$ and thus finite
superconformal chiral gauge theories in four dimensions can be constructed
\refs{\ks,\vafa,\hu}. Threebrane charge on the orbifold singularity will be enhanced by the
orbifold projection by a factor of $|\Gamma_G|$ and it takes values at K$({\bf R}^6/\Gamma_G)$
with compact support or K$({\bf S}^6/\Gamma_G)$.  Gauge and chiral
multiplets are determined by the orbifold projection of the spectrum into $\Gamma_G$
invariant states. Thus the supersymmetry, spectra and interactions of the world-volume
theory depends on the choice of the group $\Gamma_G$. Similar descriptions are applied to
theories in two-dimensions on the worldsheet of Type IIB D1-branes in orbifold
singularities of the type ${\bf C}^4 /\Gamma_G$ \refs{\mohri,\angel}. Here the choice of
the $\Gamma_G$ corresponds with different enhanced worldsheet supersymmetries.

In general the brane charge of these systems take values at K$({\bf
R}^{p+1}/\Gamma_G)$ with compact support or K$({\bf S}^{p+1}/\Gamma_G)$. Using equivariant
Bott periodicity for complex
groups one immediately find that these groups are non-zero for odd values of $p$ and more precisely for
$p=3,5,7.$ Moreover just as in the
previous case for the fivebrane, the additional structure of the $\Gamma_G$ action of the
equivariant K-theory provides the necessary tools to construct sensible gauge theories on
the world-volume of D-branes.

D-brane theory in orbifold singularities might be generalized to more general singularities
such as conifolds. In the context of AdS/CFT correspondence the field theory of branes in conifolds
was originally discussed in \kw\ (and
further studied for singular spaces at \conifold),
where the ${\bf S}^5/\Gamma_G$ is substituted by ${\rm SU(2)}\times {\rm
SU(2)/U(1)}$ and in general by an Einstein manifold $X$. It would be interesting to
pursue an appropriate description of the brane charges and their
classification of supersymmetric D3-branes (coming from $9-\overline{9}$ pairs) in conifold
singularities in terms of
an suitable generalization of the equivariant K-theory.

\vskip 2truecm

\newsec{Brane Charge and Equivariant K-theory}

In this section we will discuss equivariant K-theory description of
configurations of Type II D-branes in abelian ALE singularities. We give a formula which generalizes the
formula to compute the RR charge found in \refs{\cheung,\moore}.
First of all let us recall the Minasian and Moore result \moore\ concerning the basic
formula for obtaining the RR charges is given by

\eqn\charge{ Q = ch(f_!E) \sqrt{\hat{A}(TX)}}
where $E$ is a U$(N)$ Chan-Paton bundle over a  $(p+1)$-dimensional submanifold $W$ of the spacetime $X$, $f_!$ is a
K-theoretic Gysin map: $f_!: K(W) \to K(X)$, $\hat{A}(TX)$ is the Dirac genus of $X$ and
$ch: K(X) \to H^*(X)$ is the usual Chern character. Formula  (5.1) was obtained from
a matching between the anomaly contribution from the chiral fermions on the world-volume $W$
given by

\eqn\action{ I = \int_{W} c \wedge Y}
and the inflow anomaly from the bulk in such a way that the ten-dimensional theory is anomaly free
\refs{\ghm,\moore}. With the definition $ch(E) = {\rm tr}_N (e^{{\cal F}/2\pi})$ it was shown in \ghm\
that $Y$ at
\action\ is given by $Y= ch(E) j^* \sqrt{\hat{A}(TX)}$ where $j: W \hookrightarrow X$ is an embedding.

At the presence of orbifold singularities ${\bf R}^{10-p}/\Gamma_G$, with $\Gamma_G$ an discrete and abelian
subgroup of some Lie group, the above action is modified as \dm\

$$I_W= \int_{W} c \wedge \tilde{Y}$$
\eqn\action{ = \int_{W} c \wedge {\rm Tr}\big( \gamma(g) e^{{\cal F}/2\pi} \big)}
where $\gamma(g)$ is an element of the discrete group $\Gamma_G$ and ${\rm Tr}$ is given by a product of
traces of
$tr_R(\gamma(g))$ and  ${\rm tr}_N (e^{{\cal F}/2\pi})$.

This action is valid for the case of flat
spaces but is has to be modified for submanifolds $W$ with non-trivial normal  bundle \moore. The purpose of this
section is to get such a modification. This formula as we shall see generalizes \charge\ and it is described
appropriately by equivariant K-theory.

To be specific consider $W$ to be a six-dimensional cycle where $N$ fivebranes can be wrapped. Assume
that these fivebranes
are placed in an ALE singularity of the ${\bf A}$-type and thus $\Gamma_G={\bf Z}_N$.

In order to see the relation with the K-theory group
K$_{\Gamma_G}(X)$ we consider the action of $\Gamma_G$ on the spacetime $X=W \times {\bf R}^4$.
For the purpose of studying fivebranes sitting in ALE  singularity in the transverse
directions of the fivebrane world-volume, the action of $\Gamma_G$ on $W$ is trivial, acting only on
${\bf R}^4$ to get $X=W \times {\bf R}^4/\Gamma_G$.
From the general theory of equivariant K-theory one have that $W$ becomes a
$\Gamma_G$-trivial space \refs{\as,\segal,\greenless}. Thus Chan-Paton bundle becomes a $\Gamma_G$-bundle and
it is given by

\eqn\trivial{ E = \bigoplus_{\mu=0}^r E_\mu \otimes \underline{ R}_\mu}
where $E_\mu = Hom(\underline{R}_\mu,E)$ and
$\underline{R}_\mu$ is the trivial bundle $W \times R_\mu$ with $\{R_\mu\}$ the set of irreducible regular
representations of $\Gamma_G$.
In this orbifold singularity the gauge group is broken from U$(N)$ to
$\prod_\mu {\rm U}(Nn_\mu)$.
Thus Chan-Paton bundle are obviously modified to be

\eqn\bunduno{ \prod_{\mu=0}^r {\rm U}(Nn_\mu) \to E \to W.}

Is well known that for the
case of trivial actions, the group K$_{\Gamma_G}(W)$ decomposes just as the tensor product
K$(W) \otimes R(\Gamma_G)$ where $R(\Gamma_G)$ is the ring of regular irreducible representations
of $\Gamma_G$ \refs{\as,\segal,\greenless} .

On the other hand chiral fermions in intersections of two cycles $Z=W_1\cap W_2$ leads to the well known
world-volume gauge anomaly ${\cal A}$ \refs{\ghm,\moore}. The only
modification is in the Chern character

\eqn\chern{ {\cal A}= ch_{\Gamma_G}(E_1) ch_{\Gamma_G}(E_2) \hat{A}(TZ)}
where $E_1$ and $E_2$ are the Chan-Paton bundles over $W_1$ and $W_2$ respectively and
$ch_{\Gamma_G}(E)$ is the equivariant Chern character given by

\eqn\bagel{ch_{\Gamma_G}(E)= {\rm Tr}\big( \gamma(g) e^{{\cal F}/2\pi}\big).}

If one rewrites the last equation in terms of properties of $W$ and $X$ only as in \moore\ this leads
to the modification of $\tilde{Y}$ in Eq. (5.3) as

\eqn\mod{ \tilde{Y} \to \tilde{Y}' = \tilde{Y} e^{{1\over 2}d}  {\hat{A}(TW)\over j^* \hat{A}(TX)}.}

Thus Eq. \action\ should to be rewritten as

\eqn\act{ I_W = \int_W c \wedge ch_{\Gamma_G}(E)\hat{A}(TW) \cdot {1 \over j^*\sqrt{\hat{A}(TX)}}.}

Following the same procedure as \moore\ we lead finally to the definition of brane charge

\eqn\firstcharge{ Q = \int_C j_* \bigg(ch_{\Gamma_G}(E)\hat{A}(TW)\bigg)
 \cdot {1 \over j^*\sqrt{\hat{A}(TX)}}}

In the equivariant case it is still valid Atiyah-Hirzebruch theorem \refs{\as,\segal} and we get

\eqn\ah{ j_*\bigg( ch_{\Gamma_G}(E)\hat{A}(TW)\bigg) = ch_{\Gamma_G}(j_!E)\hat{A}(TX)}
where $j_!: {\rm K}_{\Gamma_G}(W) \to {\rm K}_{\Gamma_G}(X)$ is a $\Gamma_G$-isomorphism
induced by the embedding $j:W \hookrightarrow X$. Thus the formula for the brane charge reads

\eqn\secondcharge{ Q= ch_{\Gamma_G}(j_!E)\sqrt{\hat{A}(TX)}}
where $ch_{\Gamma_G}: {\rm K}_{\Gamma_G}(X) \to H^*(X;{\bf Q}) \otimes R(\Gamma_G).$
One can further proceed specifying $\Gamma_G={\bf Z}_N$ and the fact that
for each $\gamma(g)\in \Gamma_G$ we have the regular irreducible representation $R_\mu$ and for each
regular representation we have a character $\chi: R(\Gamma_G) \to {\bf Z}$ given by $\chi_\mu
\equiv tr_R({R_\mu}(g))$. As we have seen $E$ is of the form given by Eq. \trivial, thus
equivariant Chern character is given by

\eqn\equivchern{ ch_{\Gamma_G}(E) = ch_{\Gamma_G}\big(\bigoplus_{\mu=0}^r E_\mu \otimes
\underline{R}_\mu\big).}
Using the fact that $ch_{\Gamma_G}$ is a ring homomorphism between K$_{\Gamma_G}(W)$ and
$H^*(W;{\bf Q})\otimes R(\Gamma_G)$ we get
\eqn\result{  ch_{\Gamma_G}(E)= \sum_{\mu =0}^r ch(E_\mu) \chi_\mu}
where $\chi_\mu= \chi({R_\mu})$ and $ch(E_\mu) = {\rm tr}_N(e^{{\cal F}_\mu/2\pi})$.
Finally, action of Gysin map on the bundle $j_!E\in K_{\Gamma_G}(W)$ is

\eqn\gysin{ j_!\big(\bigoplus_{\mu=0}^r E_\mu \otimes R_\mu\big)= \bigoplus_{\mu=0}^r j_!(E_\mu)
\otimes R_\mu. }
Thus the formula for the brane charge in a ${\bf C}^2/\Gamma_G$ orbifold singularity
says that it takes values in the equivariant K-theory group K$(X) \otimes R(\Gamma_G)$

\eqn\forcharge{ Q = \sum_{\mu=0}^r ch(j_!E_\mu) \chi_\mu \sqrt{\hat{A}(X)}.}
This formula can be  generalized for non-trivial action and thus obtaining  that the
brane charge can take values at K$_{\Gamma_G}(X).$

Although we have computed the brane charge for ALE singularities, this procedure immediately
generalizes to other singularities as ${\bf C}^k/\Gamma_G$ with $k=3,4$, {\it i.e.} for
threefolds and fourfolds singularities.

\vskip 2truecm

\newsec{Concluding Remarks}
In this paper we have further investigated (non-)BPS states on orbifold singularities using for that
the language of equivariant K-theory. The idea of describing pairs of D-branes in orbifold
singularities in terms of equivariant K-theory was originally argued by Witten in \witten.
Mathematically
the relation between the ordinary K-theory and the equivariant ones is the additional action
of a group on the spacetime $X$ and the relevant Chan-Paton bundles. We have take
this group action to be a discrete and abelian group action of
$\Gamma_G$ on the spacetime $X$. We have shown that this mathematical
framework describes and classifies correctly configurations of (non-)BPS branes in orbifold
singularities. We have seen that several results of 9-branes in orbifold singularities are
nicely reproduced in the K-theory formalism. Among them of particular interest is the result
of \julie, where inflow mechanism for branes on singularities is shown to be play an important role.
 We have also discussed
how the incorporation of lower-dimensional branes than 9, follows a parallel treatment given in
\witten, in the language of equivariant K-theory. In addition in discussing  examples
 we consider Type I zero
and -1 branes on specific orbifold singularities. Moreover for Type I theory since Bott periodicity gives
a correspondence between 0-branes 8-branes and -1-branes an 7-branes (with period equal to 8),
the equivariant Bott periodicity determines a map between the corresponding branes and their
associated mirror images produced by the orbifold projection.

We have also discussed the description of fivebranes in orbifold singularities of the ALE type,
for Type I and
IIB superstring theories. For Type IIB theory we have placed coincident fivebranes at a point
of an ALE singularity ${\bf C}^2/\Gamma_G$ in the transverse directions of the fivebranes.
Following Witten results for the fivebrane \witten, the fivebrane charge is determined by the
equivariant K-theory group K$_{\Gamma_G}(Z_\zeta)$ of the ALE space $Z_\zeta$ coming from the
minimal resolution of the ALE
singularity ${\bf C}^2/\Gamma_G$. This system is completely determined by a pair of gauge
bundles over $Z_\zeta$. Description of these pair in terms of associated pair of vector bundles
leads naturally to identify an element of K$_{\Gamma_G}(X)$ with a pair of tautological vector bundles
arising in the Kronheimer-Nakajima construction of Yang-Mills instantons on ALE spaces
\kn. This relation is not a coincidence as we have shown using
equivariant K-theory to get all relevant topological information which enters in the description of non-trivial RG fixed points of six-dimensional
theories in ALE singularities \refs{\six,\bi}. Similarly the description of Type I fivebrane
or small instantons in ALE
singularities \refs{\six,\bi} leads to the description of the charge of the fivebranes in terms of the
equivariant real KO-theory group KO$_{\Gamma_G}(X)$. The relation between both K-theory groups
(complex and real) is a multiplication by $2 |\Gamma_G|$. We shown that
KO$_{\Gamma_G}(X)$ determines also all data of the dynamics of the non-trivial RG fixed points
of the supersymmetric gauge theories on the world-volume of the Type I fivebranes.

In section 5 we have reexamined the cohomological formula to get the RR charge of D-branes, obtained in
\refs{\moore,\cheung} in the context of branes in singularities. We have found a formula
which generalizes formula (5.1) in the context of cohomological terms of equivariant K-theory.

 One can study the theory in four dimensions of Type IIB theory
D3-branes on ${\bf R}^4 \times {\bf C}^3/\Gamma_G$ where $\Gamma_G$ being a discrete abelian or non-abelian
subgroup of SU$(K)$ with $K=2,3,4$. Also one could study the two-dimensional wold-sheet theory on coincident
D1-branes on orbifolds ${\bf C}^4/\Gamma_G$, where $\Gamma_G$ is a discrete and abelian (or non-abelian)
subgroup of SU(4) \mohri. It would be interesting a better understanding of such theories in
the context of equivariant K-theory describing (non-)BPS brane configurations. Finally AdS/CFT
correspondence on orbifolds $AdS_k \times {\bf S}^{10-k}/\Gamma_G$ would be an intersting arena
to study non-BPS brane configurations and it is tantalizing to speculate that equivariant
K-theory may be an invaluable tool to study these configurations within
AdS/CFT correspondence. The problem of finding an appropriate mathematical framework to generalize
to $[H]\not= 0$ argued in \witten\ remains still open in our present description of branes in orbifolds.

Very recently Horava in \horava, has studied non-stable brane-antibrane systems in Type IIA theory
and its relation to Matrix theory. Specifically he shown that all supersymmetric Type IIA D-branes
can be seen as bound states of non-supersymmetric 9-branes. It would be extremely interesting to apply
Horava results and the results of the present paper to study Type IIA D-branes and possibly to
M-theory membranes and fivebranes in orbifolds following
\orbi.

\vskip 2truecm


\centerline{\bf Acknowledgements}

I would like to thank A. G\"uijosa for many useful comments and suggestions.
I am grateful to A.M. Uranga for stimulating discussions and a critical review of the
manuscript.
It is a pleasure to thank E.~Witten for his hospitality in the Institute for Advanced
Study. This work
is supported by a Postdoctoral CONACyT fellowship (M\'exico) under the
program {\it Programa de Posdoctorantes: Estancias Posdoctorales
en el Extranjero para Graduados en Instituciones Nacionales 1997-1998}.

\vskip 2truecm


\listrefs

\end